\documentclass[twocolumn,showpacs,prl]{revtex4}
\usepackage{amsmath,amssymb,graphicx,color}

\newcommand{\MQSS}{M_{\text{QSS}}}

\begin{document}

\title{Out--of--equilibrium tricritical point in a system
  with long-range interactions}
\author{Andrea Antoniazzi$^{1},$%\thanks{andrea.antoniazzi@unifi.it},
Duccio Fanelli$^{1,2}$,%\thanks{Duccio.Fanelli@manchester.ac.uk}
Stefano Ruffo$^{1}$,%\thanks{stefano.ruffo@unifi.it},
Yoshiyuki Y. Yamaguchi$^{3}$}%\thanks{yyama@amp.i.kyoto-u.ac.jp}
\affiliation{ 1. Dipartimento di Energetica and CSDC, Universit\`a di
  Firenze, and INFN, via S. Marta, 3, 50139 Firenze, Italy\\
  2. Theoretical Physics, School of Physics and Astronomy, University of Manchester,Manchester M13 9PL, 
  United Kingdom \\
  3. Department of Applied Mathematics
  and Physics, Graduate School of Informatics, Kyoto University,
  606-8501, Kyoto, Japan}
\date{\today}

\begin{abstract}
Systems with long-range interactions display a short-time relaxation towards Quasi Stationary States (QSSs) whose
lifetime increases with system size. With reference to the Hamiltonian Mean Field (HMF) model,
we here show that a maximum entropy principle, based on Lynden-Bell's pioneering idea of ``violent relaxation", 
predicts the presence of out--of--equilibrium phase transitions separating the relaxation towards
homogeneous (zero magnetization) or inhomogeneous (non zero magnetization) QSSs. When varying the initial 
condition within a family of ``water-bags" with
different initial magnetization and energy, first and second order phase transition lines are 
found that merge at an out--of--equilibrium tricritical point. Metastability is theoretically 
predicted and numerically checked around the first-order phase transition line.
\end{abstract}


\maketitle

The emergence of phase transitions in thermal equilibrium  
is a well understood and widely studied phenomenon. 
For {\it short-range} interactions, phase transitions have been
explained in the context of {\it equilibrium} statistical mechanics.
Analytically, they are signalled by the appearance of 
singularities in the thermodynamic potentials at specific points (or regions) of
the control parameter space (temperature, energy, external magnetic field, etc.).
The situation becomes more intricate when one considers systems with {\it long-range} 
interactions \cite{Houches02}. 
In this case, the property of {\it additivity}, which is used when deriving the canonical
ensemble from the microcanonical, is no more valid.
Due to this intrinsic difficulty, it has been only in the last decade
that phase transitions have been analyzed with reference to models with long-range interactions, 
revealing a rich variety of interesting situations~\cite{BarreBouchet}.
For instance, the inequivalence of microcanonical and canonical ensembles requires a
separate analysis of the phase diagram in the two ensembles. Phase transitions of first and second
order are found, with related tricritical points, but their location in the control
parameters space is not the same in the two ensembles~\cite{BarreMukamel}. This also
justifies why one can find negative specific heat in the microcanonical ensemble~\cite{LyndenBell68}.

Moreover, focusing on dynamical aspects, remarkable out-of-equilibrium features are displayed for 
long-range systems. It is for instance well known that   
such systems get trapped in long-lasting Quasi-Stationary-States (QSSs)~\cite{Latora}, before 
relaxing to thermal equilibrium. The existence of QSSs was recognized in a cosmological setting 
(see~\cite{Konishi} and references therein) and subsequently re-discovered in other
contexts, e.g. plasma-wave interactions \cite{ElskensEscande}. Importantly, when performing the limit 
$N \rightarrow \infty$ (where $N$ is the number of particles) {\it before} the infinite time limit, the system
remains permanently confined in QSSs. Consequently, QSSs represent  
the solely experimentally accessible dynamical regimes for systems composed by a
large number of particles subject to long-range couplings. This includes physical systems
of paramount importance, ranging from Free Electron Lasers~\cite{barre} to 
ion and particle beams~\cite{turchetti}. The emergence of QSSs has originated an intense 
debate on the foundation of statistical mechanics \cite{Houches02}: surprisingly, the QSSs 
keep memory of the initial condition and, 
consequently, they cannot be interpreted by resorting to traditional Boltzmann-Gibbs treatments. In a recent series 
of papers \cite{barre, antoniazzi-06, califano}, an approximate analytical theory based on the Vlasov equation 
and inspired by the pioneering work of Lynden-Bell~\cite{LyndenBell68} has been elaborated. 
This is a fully predictive approach that enables one to
explain the appearence of QSS from first principles. 

\begin{figure}[htbp]
  \centering
%  \vspace*{1em}
  \includegraphics[width=6cm]{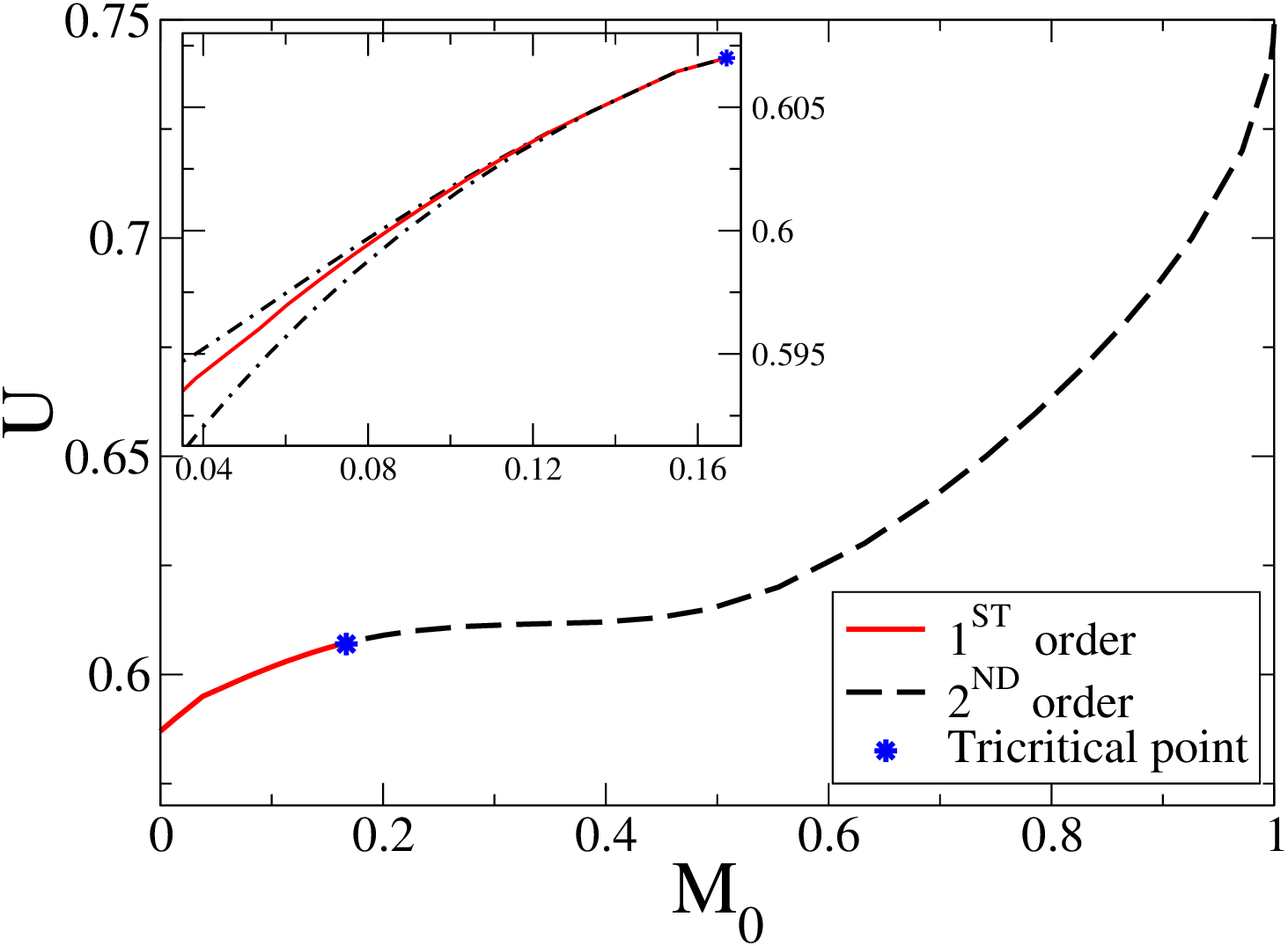}
  \caption{Theoretical phase diagram on the control parameter plane $(M_{0},U)$: second order phase transition line (dashed); 
  first order phase transition line (full); tricritical point (full dot). Inset: magnification of the first order 
  phase transition region and limits of the metastability region (dash-dotted).
} 
  \label{fig:phase-diagram}
\end{figure}

In this Letter we shall take one step forward and demonstrate, both analytically and numerically,
that out--of--equilibrium phase transitions occur in the QSSs, separating qualitatively different dynamical regimes. 
The analysis is 
carried out for the Hamiltonian Mean Field (HMF) model~\cite{antoni-95}, which describes 
the motion of $N$ coupled rotators. We determine the out--of--equilibrium phase diagram for a family
of initial ``water-bag" distributions, displaying first and second order 
transition lines between homogeneous (zero magnetization) and inhomogeneous (non zero magnetization) QSSs, 
that merge together at a {\it tricritical point}. The metastability region near the first order phase 
transition line is also studied.

The HMF model has the following Hamiltonian
\begin{equation}
\label{eq:ham}
H = \frac{1}{2} \sum_{j=1}^N p_j^2 + \frac{1}{2 N} \sum_{i,j=1}^N [1-\cos(\theta_j-\theta_i)]~,
\end{equation}
where $\theta_j$ represents the orientation of the $j$-th rotator
and $p_j$ stands for the conjugated momentum. To monitor the evolution of
the system, it is customary to introduce the magnetization, 
an order parameter defined as $M=|{\mathbf M}|=|\sum {\mathbf m_i}|
/N$, where ${\mathbf m_i}=(\cos \theta_i,\sin \theta_i)$ is the
magnetization vector. The HMF model shares many similarities with 
gravitational and charged sheet models~\cite{Konishi,ElskensEscande} and has been extensively studied 
\cite{chavanis} as a paradigmatic representative of the broad class of systems with long-range interactions.
Equilibrium statistical mechanics calculations~\cite{antoni-95} reveal the existence of  a second-order 
phase transition at the critical energy density $U_{c}=3/4$: below this threshold value the Boltzmann-Gibbs
equilibrium state is magnetized.
 
As previously reported \cite{antoni-95,Latora}, starting from some {\it out--of--equilibrium} initial
conditions, for energies below $U_c$, the system gets trapped in QSSs,
whose lifetime diverges when increasing the number $N$ of rotators. 
In this regime, the magnetization is lower than predicted by the 
Boltzmann--Gibbs equilibrium and the system displays non Gaussian velocity
distributions~\cite{tsallis,barre-06,antoniazzi-06}.

In the limit of $N \rightarrow \infty$, the system is described by
the following Vlasov equation
\begin{equation}
\frac{\partial f}{\partial t} + p\frac{\partial f}
{\partial \theta} - \left( M_x[f] \sin \theta - M_y[f]\cos \theta \right) \frac{\partial f}{\partial p}=0,
\label{eq:VlasovHMF}
\end{equation}
where $f(\theta,p,t)$ is the one-body microscopic distribution function, $M_x[f]=\int f \cos \theta d\theta dp$ and 
$M_y[f]=\int f \sin \theta d\theta dp$.     
Hereafter, invoking rotational symmetry, we shall assume $M_{y}=0$ and denote $M_x$ as $M$. 
With reference to cosmological applications, Lynden-Bell~\cite{LyndenBell68} proposed an analytical approach
to describe the stationary solutions of the Vlasov equation. He considered 
the coarse-grained distribution function $\bar{f}$ over a finite grid and
associated an entropy $s[\bar{f}]$ to such distribution. The statistical equilibrium obtained by maximizing such entropy, 
while imposing the conservation of Vlasov dynamical invariants, would determine the initial ``violent" relaxation.
This idea was later applied to the two-dimensional Euler equation~\cite{Chavanis96}.

Consider now a family of water-bag initial distributions,  
which take a constant value $f_0$ inside the phase-space domain $D$ specified by
\begin{equation}
D = \{(\theta,p)\in [-\pi,\pi] \times [-\infty,\infty]~|~ |\theta|<\Delta\theta, ~|p|<\Delta p\},
\end{equation}
where $0\leq\Delta\theta\leq\pi$ and $\Delta p \geq 0$.
The normalization condition fixes $f_0=1/(4\Delta\theta\Delta p)$.
Hence, the initial magnetization $M_{0}$ and the energy density $U$ can be expressed as functions
of $\Delta\theta$ and $\Delta p$
\begin{displaymath}
M_{0} = \dfrac{\sin(\Delta\theta)}{\Delta\theta} ,
\quad
U = \dfrac{(\Delta p)^{2}}{6} + \dfrac{1-(M_{0})^{2}}{2}~,
\end{displaymath}
which in turn implies that the initial water-bag profiles are uniquely determined by $M_{0}$ and $U$,
which take values in the ranges $0 \leq M_{0} \leq 1$ and $U \geq (1 - M_0^2)/2$.
With reference to this specific case, the Lynden-Bell entropy constructed from the coarse-grained 
function $\bar{f}$ reads
\begin{equation}
s[\bar{f}]=-\int \!\!{\mathrm d}p{\mathrm d}\theta \, 
\left[\frac{\bar{f}}{f_0} \ln \frac{\bar{f}}{f_0}
+\left(1-\frac{\bar{f}}{f_0}\right)\ln
\left(1-\frac{\bar{f}}{f_0}\right)\right].
\label{eq:entropieVlasov}
\end{equation}
Requiring that this entropy is stationary, we obtain the following distribution~\cite{antoniazzi-06} 
\begin{equation}
  \label{eq:barf}
  \bar{f}_{\text{QSS}}(\theta,p)=
  \frac{f_0}{e^{\beta (p^2/2 -M[\bar{f}_{\text{QSS}}]\cos\theta)+\lambda p+\alpha}+1},
\end{equation}
where $\beta$, $\lambda$ and $\alpha$ are Lagrange multipliers associated with 
the conservation of energy, momentum and mass. The magnetization in the QSS,
$\MQSS=M[\bar{f}_{\text{QSS}}]$, and the values of the multipliers 
are obtained by solving the self-consistent equations which follow by imposing the 
conservation laws mentioned above. Since we look for solutions where the total momentum is zero, 
the Lagrange multiplier $\lambda$ vanishes. It should also be emphasized that multiple local maxima
of the entropy are in principle present when solving the variational problem. 

Let us introduce the control parameter plane $(M_{0},U)$ (these are indeed the analogues
of thermodynamic fields in equilibrium). In Fig. \ref{fig:phase-diagram} we
plot the transition line that divides the region of the plane where the global maximum of Lynden-Bell
entropy has $\MQSS=0$ (homogeneous state), where $\bar{f}_{\text{QSS}}(\theta,p)$ does not
depend on $\theta$, from that where the maximum is for $\MQSS>0$ (inhomogeneous state).  
This means that, e.g., when fixing the initial magnetization $M_0$ and decreasing the energy density $U$, 
the system undergoes an out--of--equilibrium phase transition from a homogeneous to an inhomogeneous state.
Along the transition line two distinct regions can be isolated: 
the dashed line corresponds to a second order phase transition, the full line refers to a first order phase transition.
First and second transition lines merge together at a tricritical point, approximately located at
$(M_{0},U)=(0.17,0.61)$. Tricritical points are a well known feature for systems at equilibrium, 
and are here shown to occur also out--of--equilibrium.

Two important remarks are mandatory at this point. 
First, contrary to the usual equilibrium treatment, we here describe the behavior of the
system at short time, when it attains a QSS. For equilibrium phase transitions, one instead
looks at the behavior of the system at long times, when the magnetization 
corresponds to a global maximum of the Boltzmann entropy (rather than 
Lynden-Bell's entropy). Second, while $U$ is a 
standard control parameter, used also for equilibrium phase transitions, the initial
magnetization $M_0$ does not appear in the standard treatment of equilibrium phase transitions.
Indeed, when using Boltzmann entropy, the HMF model undergoes a second order phase
transition at $U_c=3/4$, independently of $M_0$. Let us notice that this transition energy
value appears in Fig. \ref{fig:phase-diagram} for $M_0=1$.
The lower edge of the metastability region, plotted in the inset, converges to $U=7/12$ for
$M_0 \rightarrow 0$, a value found in \cite{barre-06} to correspond to the destabilization of
the homogeneous (zero magnetization) state in the Vlasov equation.

\begin{figure}[htbp]
%\vspace{0.5cm}
%\usepackage{dcolumn}
\centering
\includegraphics[width=6cm]{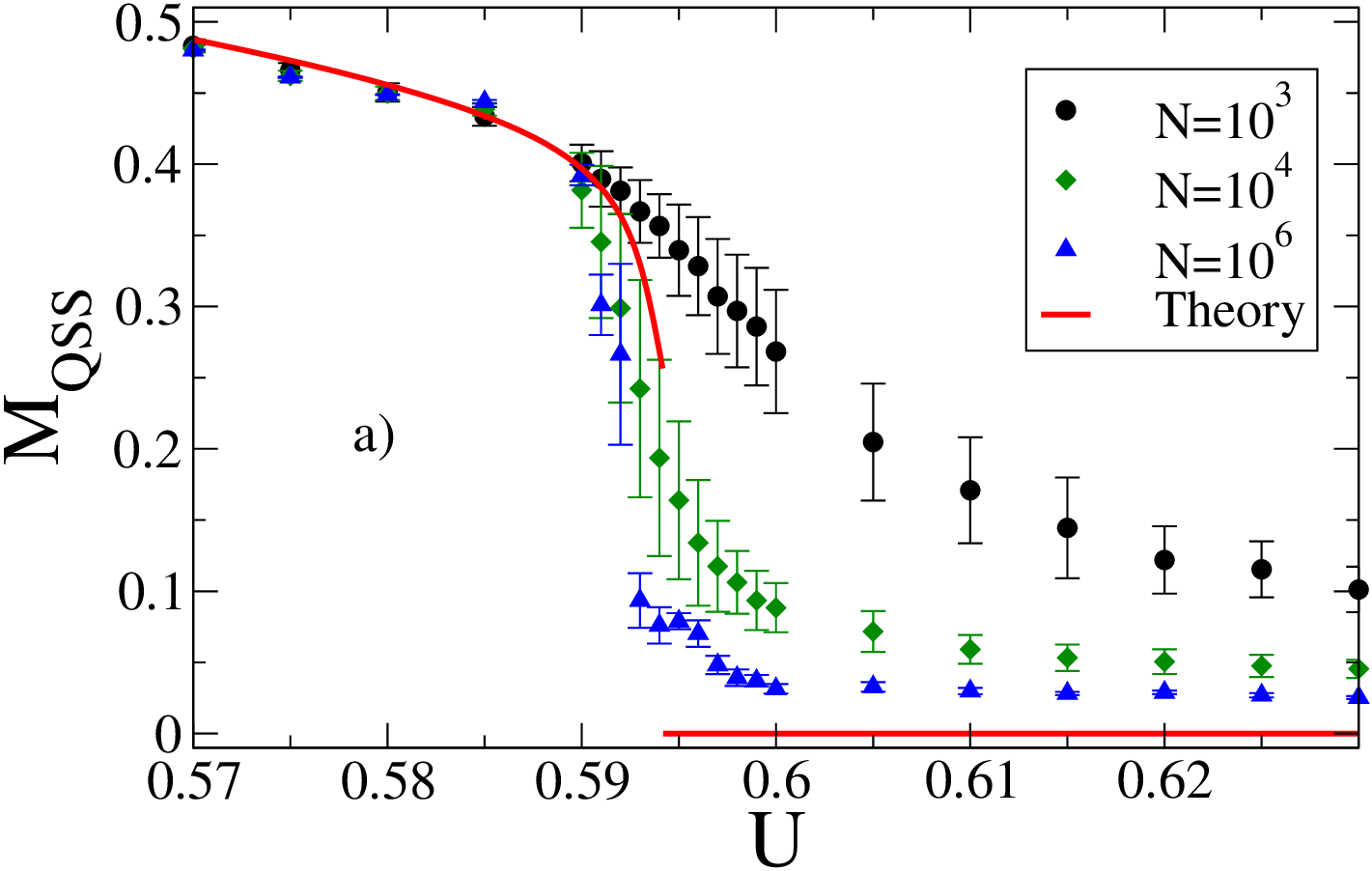}\vspace{1cm}
\includegraphics[width=6cm]{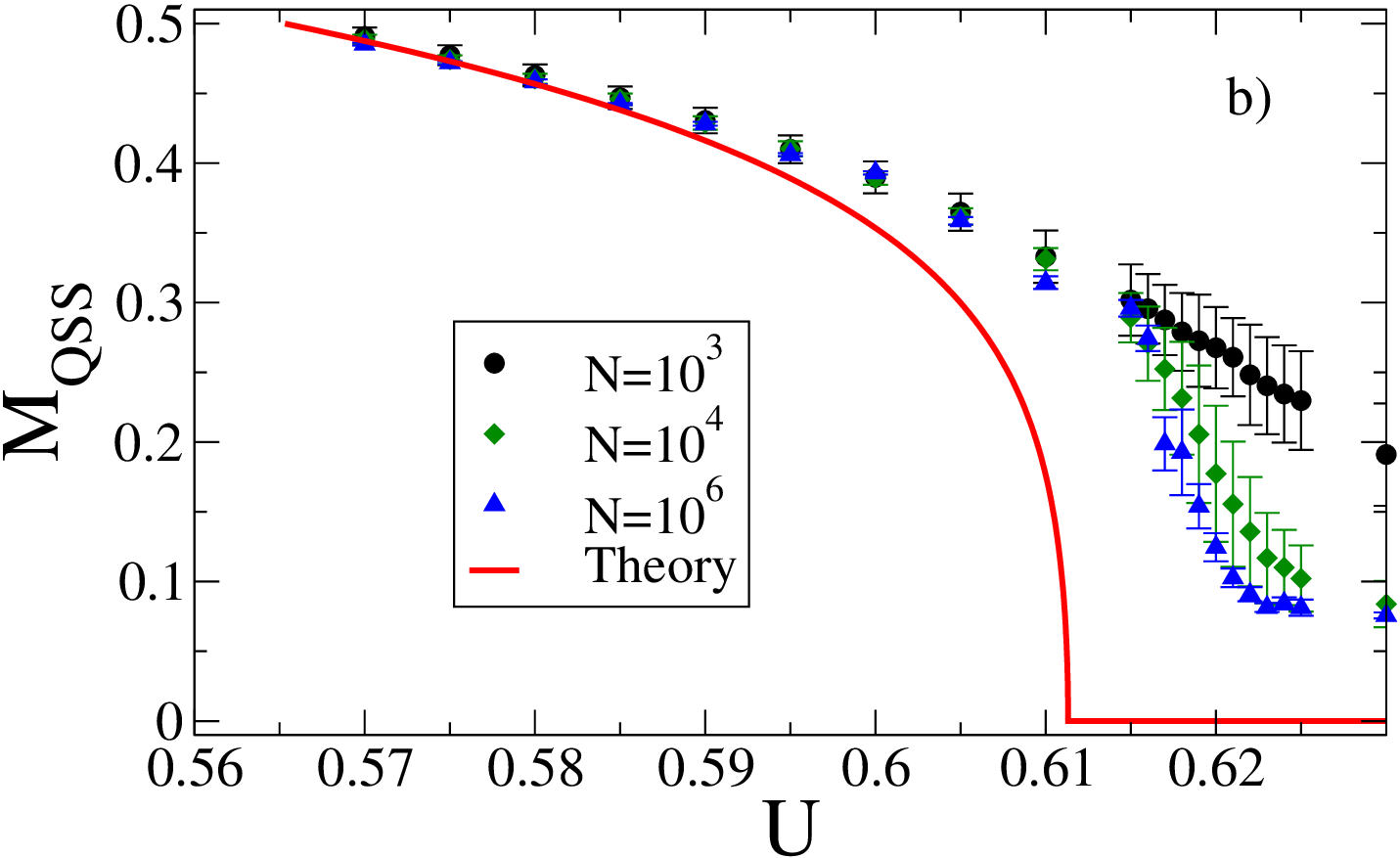}\vspace{0.5cm}
\caption{$\MQSS$ as a function of $U$ for (a) $M_{0}=0.05$ (first order phase transition). (b) 
$M_{0}=0.30$ (second order phase transition). The number of initial realizations is
$10^5$ ($N=10^3$), $10^4$ ($N=10^4$), $10^2$ ($N=10^6$) and the averaging time $20 < t \leq 100$.}
\label{fig:transition}
\end{figure}

To assess the correctness of the above theoretical picture, we have performed numerical simulations
of the HMF model (\ref{eq:ham}) for finite $N$. To extrapolate the relevant behavior occurring in the limit $N\to\infty$, 
where the Vlasov description applies, we have varied $N$ from $N=10^3$ to $N=10^6$. 
We have chosen two values of $M_0$, one in the first order phase transition region
($M_0=0.05$) and the other in the second order region ($M_0=0.3$). 
For these two values, we plot in Fig.\ref{fig:transition} $\MQSS$ versus $U$  for increasing values of $N$.
The magnetization in the QSS is determined by averaging over time ($20 < t \leq 100$).
Points and error bars in Fig.\ref{fig:transition} represent averages and standard deviations over several
different initial conditions. The result of the
theoretical analysis (full curve) is in reasonable agreement with the simulations, and the 
agreement improves, as expected, when $N$ is increased. It must be also stressed
that the predictions of the theory have no adjustable fitting parameter. 
This confirms the adequacy of Lynden-Bell's theoretical framework. 
The discrepancies detected near transition energies are discussed in 
\cite{califano} and shown to correspond to regions where Lynden-Bell's entropy is substantially
flat, which implies the existence of an extended basin of states where the system 
can possibly be trapped. 

\begin{figure}[htbp]
  \centering
  \vspace*{0.5em}
  \includegraphics[width=6cm]{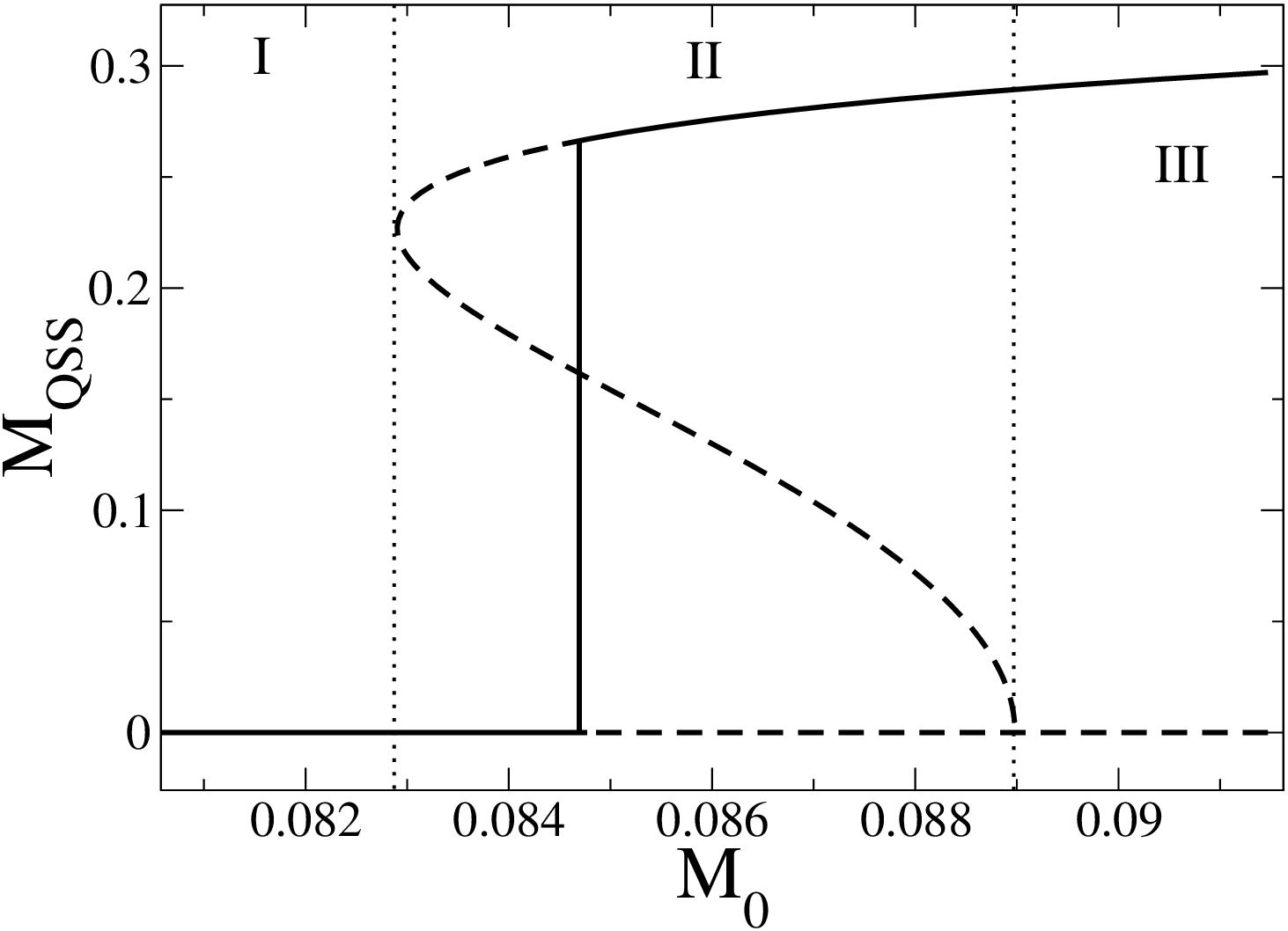}
  \caption{$\MQSS$ as a function of $M_{0}$ for $U=0.6$.}
  \label{fig:multivalue}
\end{figure}

To clarify the behavior of $\MQSS$ in the first order phase transition region,
we consider the energy value $U=0.6$ and solve the self-consistent equations for different initial 
magnetization values $M_0$.
Results are displayed in Fig.\ref{fig:multivalue}, where $M_{QSS}$ is plotted as a function of $M_0$.
An inspection of this figure suggests to identify three regions, 
delimited by different values of the control parameter $M_{0}$. For $M_{0}\lesssim 0.083$ (region I),
only one solution of the self-consistent equations is found which corresponds to an entropy maximum and is associated 
with a homogeneous QSS ($\MQSS=0$).
For $M_{0}\gtrsim 0.089$ (region III), two solutions are instead detected: the one with $\MQSS>0$ is stable,
while that with $\MQSS=0$  is unstable.
Finally, for $0.083\lesssim M_{0}\lesssim 0.089$  (region II), three solutions are obtained: two of them 
(respectively with $\MQSS^1=0$ and $\MQSS^3 \neq 0$) are local maxima of the entropy, thus thermodynamically stable.
The third one ($\MQSS^2$, with $\MQSS^1 < \MQSS^2 < \MQSS^3$) is a local minimum, hence
unstable. The expected equilibrium state is consequently determined by evaluating the coarse-grained entropy in 
correspondence of the former two stationary points and deducing the actual global maximum. Aiming at fully resolving 
the magnetization curve for $U=0.6$, one has to perform this additional test for each selected value of  
$M_0$: a direct calculation enables to track the profile outlined in Fig. \ref{fig:multivalue} with a tick solid line. 

\begin{figure}[h!]
  \centering
\includegraphics[width=7.5cm]{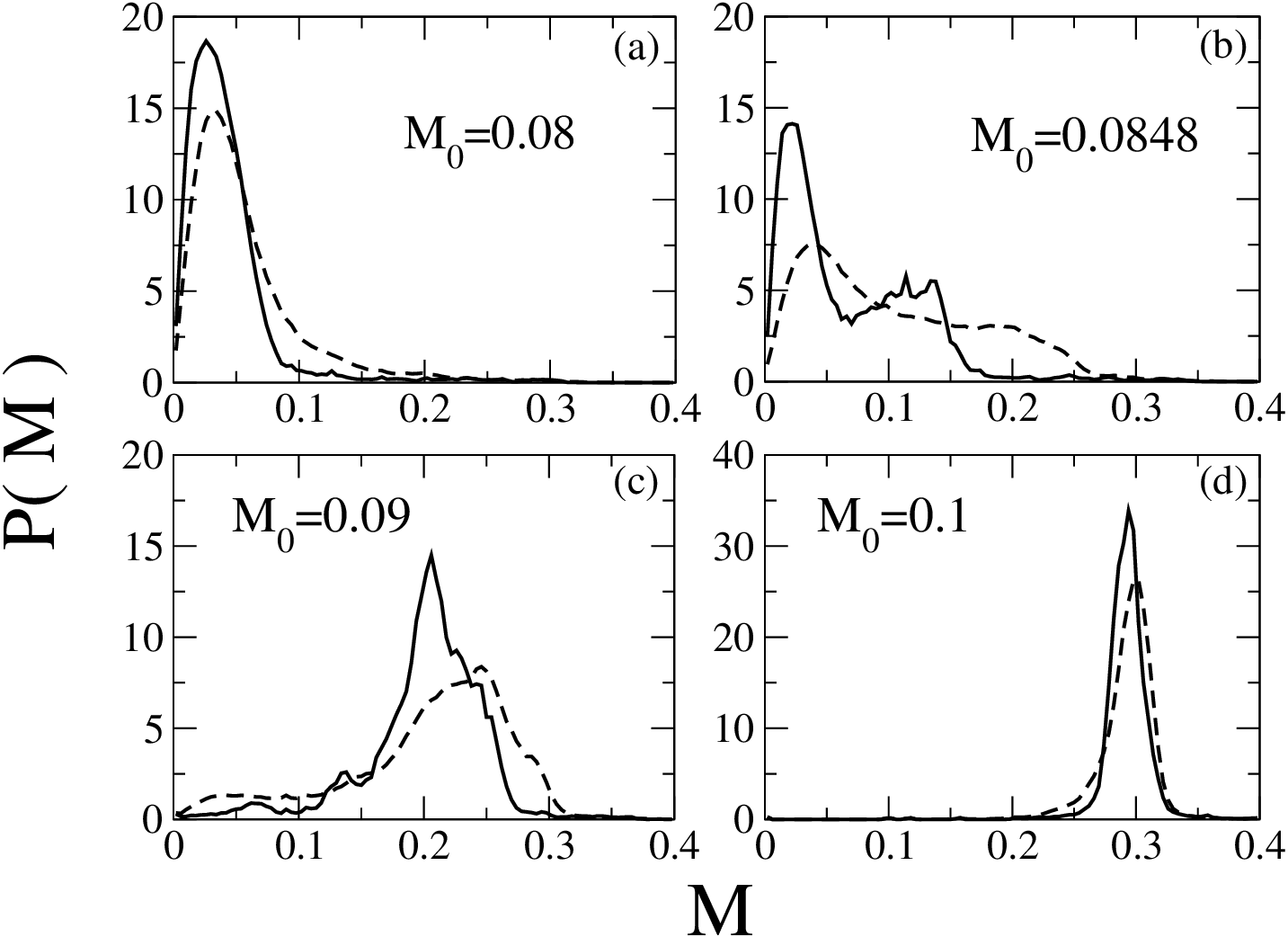}%\vspace{0.5cm}
  \caption{Probability distribution functions of $M$.
    Dashed (resp. solid) lines stand for $N=10^{5}$ (resp. $N=10^{6}$), 
    averaged over $10^{3}$ (resp. $10^{2}$) independent realizations. $M$ is always
    sampled in the time range $20 < t \leq 100$.}
  \label{fig:Mdist}
\end{figure}

Dynamically, it could however happen that the system is eventually prevented from approaching
the most probable state as predicted by the theory, when initially prepared to fall in region II. While exploring the 
phase-space, and due to metastability, the system could in fact remain indefinitely trapped in the proximity 
of the local maximum.
The edges of region II correspond to the lateral edges of the metastability region reported in the inset of 
Fig. \ref{fig:phase-diagram}.
The existence of homogeneous and inhomogeneous phases, corresponding to different local maxima
of the entropy, can be checked by computing the probability distribution function 
of $M$. In Fig.\ref{fig:Mdist} we report the histograms
of the magnetization computed in the time interval $20 < t\leq 100$ 
for $U=0.6$ and distinct values of $M_0$.  
When $M_{0}=0.08$ the system falls in region I of Fig.~\ref{fig:multivalue}:
only one peak is here observed around $M=0$, the mean value being slightly different from zero due to
finite size effects (see Fig.~\ref{fig:Mdist}(a)). For $M_{0}=0.1$, i.e. in region III,
an isolated peak is manifested, associated with an inhomogeneous state (see Fig.~\ref{fig:Mdist}(d)).   
For $M_{0}=0.0848$ (region II) two peaks are identified at $M \sim 0$ and 
$M\sim 0.1$ for $N=10^{6}$ (the peaks are shifted to the right for $N=10^5$ due again to finite
size effects), implying the existence of two local maxima of the entropy (see Fig.~\ref{fig:Mdist}(b)). The situation
in Fig.~\ref{fig:Mdist}(c) is intermediate between the single-peaked and the double-peaked distribution
although the $M_0$ value lies inside region III, possibly due to finite size effects.
One should add that, close to the transition, the positions of the magnetized peaks are only 
in rough agreement with the theory.
  \vspace*{1.5em}

\vspace*{1em}

In this Letter, we have investigated the emergence of out-of-equilibrium Quasi Stationary States (QSSs)
in the Hamiltonian Mean Field (HMF) model, a paradigmatic representative of systems with long-range 
interactions. 
We have proved the existence of  out--of--equilibrium first and second order phase transitions. 
The transition lines merge at a tricritical point.
Coexistence of homogeneous (zero magnetization) and inhomogeneous (non zero magnetization) phases is 
present at the first order phase transition line and a metastability region is revealed. 
Such transitions are expected generically in models with long-range interactions, see e.g. Ref. \cite{Johal}.
Our conclusions are analytically derived using an approach pioneered by Lynden-Bell \cite{LyndenBell68}.
The agreement with the simulations represents an a posteriori validation of Lynden-Bell's
scenario.
Besides their intrinsic theoretical relevance, we expect our results to translate into novel 
experimental solutions, with reference to those applications where long range forces are 
active and QSSs have been observed (e.g. plasmas, Coulomb systems).

\vspace*{0.5em}
\noindent
{\bf Acknowledgements}: 
YYY has been supported by the Ministry of Education, Science, Sports and Culture,
Grant-in-Aid for Young Scientists (B), 16740223, 2006.
A.A., D.F. and S.R. acknowledge financial support from the PRIN05-MIUR project {\it
Dynamics and thermodynamics of systems with long-range interactions}. S.R. thanks 
Kyoto University for hospitality and JSPS (contract No. S-06046) for financial support.

\end{document}